\documentclass[a4paper,english,prb,twocolumn]{revtex4}
\usepackage{ae} 
\usepackage[T1]{fontenc}
\usepackage[ansinew]{inputenc}
\usepackage{amsmath}
\usepackage{amssymb}
\usepackage{amsfonts}
\usepackage[dvips]{graphicx}
\usepackage{ifthen}
\usepackage{eurosym}
\usepackage{babel}

\newcommand{\ka}[1]{\langle #1\rangle}

\renewcommand{\d}{\text{d}}
\begin{document}

\title{Finite temperature quantum statistics of H$_3^+$ molecular ion}
\author{Ilkka Kyl\"anp\"a\"a and Tapio T.~Rantala}
\affiliation{Tampere University of Technology, Department of Physics,
P.O. Box 692, FI-33101 Tampere, Finland}
\date{\today}

\begin{abstract}
Full quantum statistical $NVT$ simulation of the five-particle system
H$_3^+$ has been carried out using the path integral Monte Carlo
method. Structure and energetics is evaluated as a function of
temperature up to the thermal dissociation limit.  The weakly density
dependent dissociation temperature is found to be around $4000$ K.
Contributions from the quantum dynamics and thermal motion are sorted
out by comparing differences between simulations with quantum and
classical nuclei. The essential role of the quantum description of the
protons is established.
\end{abstract}

\maketitle

\section{Introduction}
The triatomic molecular ion H$_3^+$ is a five-body system consisting
of three protons and two electrons. Being the simplest polyatomic
molecule it has been the subject of a number of theoretical and
experimental studies over the years
\cite{Oka92revmodphys,Oka03jcp,Jaquet06,Kreckel08jcp,Adamowicz09jcp1}.
Experimentally, the H$_3^+$ ion was first detected in 1911 by Thompson
\cite{Thomson11}, however, definite spectroscopic studies were carried
out not until 1980 by Oka \cite{Oka80prl}. Since then, this five-body
system has proven to be relevant, also in astrophysical studies
concerning the interstellar media and the atmosphere of gas planets.
Therefore, low-density high-temperature H$_3^+$ ion containing
atmospheres have been studied experimentally \cite{Miller08ApJ} as
well as computationally \cite{Koskinen09ApJ}.

Until now, the computational approaches have consistently aimed at
finding ever more accurate potential energy surfaces (PES) for H$_3^+$
at zero Kelvin, and consequent calculations of the rovibrational
states \cite{Meyer86jcp,Jaquet94jcp}. These calculations include
Born--Oppenheimer (BO) electronic energies in various geometries often
supplemented with adiabatic and relativistic corrections
\cite{Cencek98jcp,Komasa09jcp}. For the study of rovibrational
transitions it is desirable to have an analytical expression for the
PES, which is usually generated using Morse polynomial fits
\cite{Meyer86jcp}. Inclusion of the nonadiabatic effects, however, has
turned out to be a cumbersome task, and so far, they have not been
rigorously taken into account \cite{Adamowicz09jcp1}.

In this work, we evaluate the full five-body quantum statistics
of the H$_3^+$ ion in a stationary state at temperatures below
the thermal dissociation at about $4000$ K. We use
the path integral Monte Carlo (PIMC) approach, which allows
us to include the Coulomb correlations
between the particles exactly in a transparent way.
Thus, we are able to monitor the fully nonadiabatic correlated
quantum distributions of particles and related energies as
a function of temperature.  Furthermore, we are able to model
the nuclei as classical mass points, in thermal motion or fixed
as conventionally in quantum chemistry, and find the difference
between these and the quantum delocalized nuclei.

The PIMC method is computationally expensive, but within the
chosen models and numerical approximations it has been
proven to be useful with exact correlations and finite temperature
\cite{Li87,Ceperley95,Pierce99,Kwon99,Knoll00,
Cuervo06,Kylanpaa07,Kylanpaa09pra}. For zero Kelvin data
with benchmark accuracies,
however, the conventional quantum chemistry or other
Monte Carlo methods, such as the diffusion Monte Carlo
\cite{Anderson92jcp}, are more appropriate.
Thus, it should be emphasized that we do not aim at competing
in precision or number of decimals with the other
approaches.  Instead, we will concentrate on physical phenomena
behind the finite-temperature quantum statistics.

Next, we will briefly describe the basics of the PIMC method
and the model we use for the ion. In the
results and discussion section we first compare our
$160$ K PIMC "ground state" to the zero Kelvin ground state,
and then, consider the higher temperature effects.

\section{Method}

According to the Feynman formulation of the quantum statistical
mechanics \cite{Fey72} the partition function for interacting
distinguishable particles is given by the trace of the density matrix:
\begin{align}
Z
= \text{Tr}~\hat{\rho}(\beta)
= \int \d R_{0}\d R_{1} \ldots
\d R_{M-1} \prod_{i = 0}^{M-1}e^{-S(R_{i},R_{i+1};\tau)},\nonumber
\end{align}
where $\hat{\rho}(\beta) = e^{-\beta\hat{H}}$, $S$ is the action,
$\beta = 1/k_{\text{B}}T$, $\tau = \beta/M$, $R_{M}=R_{0}$ and $M$ is
called the Trotter number. In this paper, we use the pair
approximation in the action \cite{Storer68,Ceperley95} for the
Coulomb interaction of charges. Sampling in
the configuration space is carried out using the Metropolis procedure
\cite{Metro53} with bisection moves \cite{Chakravarty98}. The
total energy is calculated using the virial estimator \cite{Herman82}.

The error estimate in the PIMC scheme is commonly given in powers of
the imaginary time time-step $\tau$.\cite{Ceperley95}  Therefore, in
order to systematically determine thermal effects on the system we
have carried out all the simulations with $\tau = 0.03
E_\text{H}^{-1}$, where $E_\text{H}$ denotes the unit of Hartree.  Thus,
the temperatures and Trotter number $M$ become fixed by the relation
$T=(k_{\text{B}}M\tau)^{-1}$.

In the following we mainly use the atomic units, where the lengths,
energies and masses are given in units of the Bohr radius ($a_0$),
Hartree ($E_\text{H}$) and free electron mass ($m_e$), respectively.

The statistical standard error of the mean (SEM) with $2$SEM limits
is used as an error estimate for the observables, unless otherwise
mentioned.

\section{Models}

Two of the five particles composing the H$_3^+$ ion are electrons.
For these, we do not need to sample the exact Fermion statistics,
but it is sufficient to assign spin-up to one electron and spin-down
to the other one.  This is accurate enough, as long as the thermal
energy is well below that of the lowest electronic triplet excitation.

We do the same approximation for the three protons, too.  This is even
more safe, because the overlap of well localized nuclear wave
functions is negligible and related effects become very hard to
evaluate, anyway.  On the other hand, however, the nuclear exchange
due to the molecular rotation results in the so called zero-point
rotations.  These too contribute to energetics less than the
statistical accuracy of our simulations.  Therefore, we ignore the
difference between ortho-H$_3^+$ ($I=3/2$) and para-H$_3^+$ ($I=1/2$).
Thus, the protons are modeled as "boltzmannons" with the mass $m_p =
1.83615267248\times 10^{3}m_e$.  The higher the temperature, the
better is the Boltzmann statistics in describing the ensemble
composed of ortho- and para-H$_3^+$.

For the $NVT$ simulations we place one H$_3^+$ ion into
a cubic box with the volume of $(300 a_0)^3$ and apply
periodic boundary conditions (PBC) and minimum image principle.
This corresponds to the mass density of $\sim 1.255\times
10^{-6}~\rm{gcm}^{-3}$.  This has no essential effect at low $T$,
but at high $T$ the finite density gives rise to the molecular
recombination balancing the possible dissociation.  Within the
considered temperature range the dissociations are very
rare.

The electrons are always simulated with the full quantum dynamics.
For the nuclei, however, we use three models to trace
the quantum and thermal fluctuations, separately.
The case of full quantum dynamics of all particles we denote
by AQ (all-quantum), the mass point model of protons by CN
(classical nuclei) and the adiabatic case of fixed nuclei by BO
(Born--Oppenheimer potential energy surface).

\begin{figure}[t]
 \begin{center}
   \includegraphics{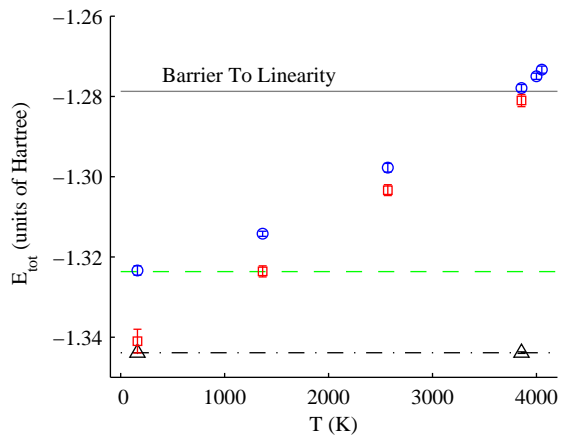}
   \caption{\label{Fig1}(Color online) Total energy of the H$_3^+$
   molecular ion as a function of temperature. Fully nonadiabatic
   quantum statistical simulations, AQ (blue circles), classical nuclei
   simulations, CN (red squares), and the equilibrium geometry
   Born--Oppenheimer simulation, BO (black
   triangles). Zero Kelvin data
   \cite{Jaquet06,Adamowicz09jcp1,Jaquet94jcp}
   is given for comparison: BO ground
   state energy at equilibrium internuclear geometry (black
   dash-dotted line), energy including the nuclear zero-point motion
   (green dashed line) and energy at the barrier to linearity (grey
   solid line). $2$SEM statistical error estimate is shown by the
   error bars from simulations at the H$_3^+$ ion density
   $(300 a_0)^3$ or $\sim 1.255\times 10^{-6}~\rm{gcm}^{-3}$.}
 \end{center}
\end{figure}

\begin{figure}[t]
 \begin{center}
   \includegraphics{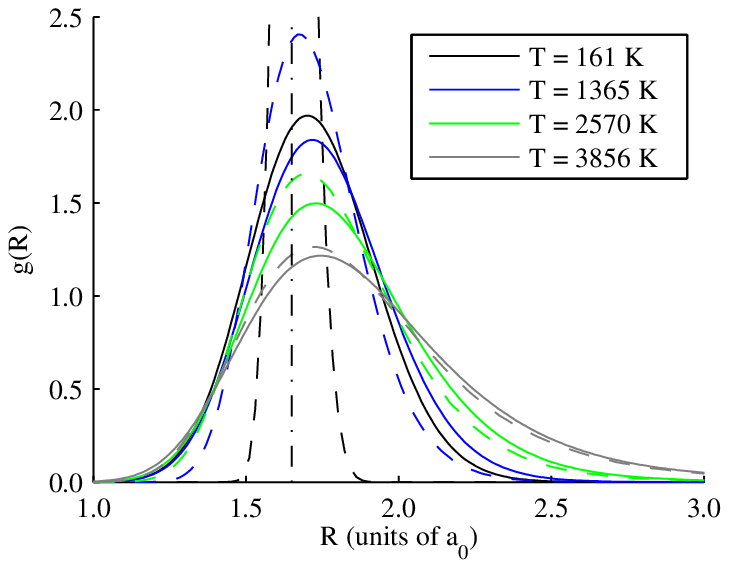}
   \caption{\label{Fig2}(Color online) Nuclear pair correlation
   functions (bond length distributions) at different temperatures
   from the quantum statistical simulations (solid lines), and from
   the classical nuclei simulations (dashed lines). The zero Kelvin
   equilibrium internuclear distance is given as a vertical black
   dash-dotted line. The distributions include the $r^2$ weight and
   normalization to unity.  (Note that the $r^2$ weight is usually not
   included in description of extended or periodic systems)}
\vskip1cm
   \includegraphics{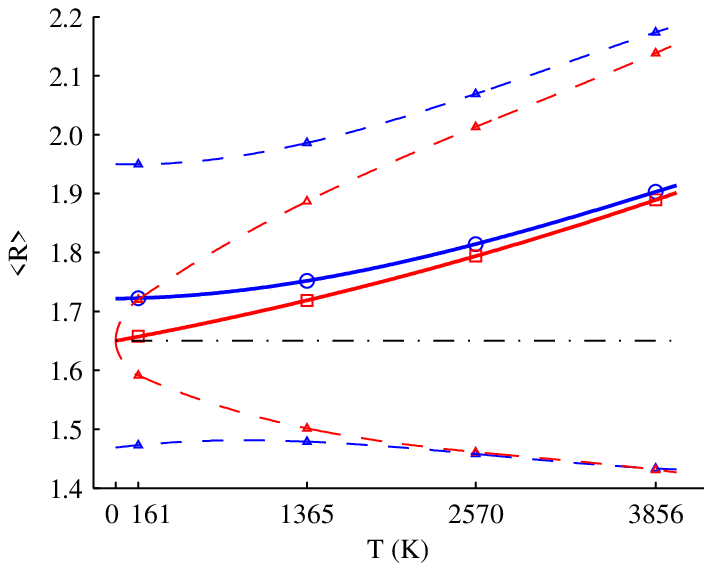}
   \caption{\label{Fig3}(Color online) Expectation values of the
      internuclear distance at different temperatures from distributions
      in Fig.~\ref{Fig2}. Quantum statistical simulations (blue
      circles) and classical nuclei simulations (red squares). The
      FWHM limits are shown by triangles (all the lines are for
      guiding the eye). The zero Kelvin equilibrium internuclear
      distance is shown as a horizontal black dash-dotted line.}
 \end{center}
\end{figure}

\section{Results and discussion}

\subsection{Ground state: zero Kelvin reference data}

The equilibrium geometry of the H$_3^+$ ion in its ground state is an
equilateral triangle $D_{3h}$ for which the internuclear equilibrium distance
is $R = 1.65 a_0$ \cite{Adamowicz09jcp1}. The best
upper bound for the electronic ground state BO energy to date is
$-1.34383562502 E_\text{H}$ \cite{Adamowicz09jcp1}. The vibrational
normal modes of H$_3^+$ are the symmetric-stretch mode $\nu_1$ and the
doubly degenerate bending mode $\nu_2$. The latter one breaks the
full symmetry of the molecule, and therefore, it is infrared active
\cite{Kreckel08jcp}.

The vibrational zero-point energy is $0.01987 E_\text{H}$, and
the so called rotational zero-point energies are $0.00029 E_\text{H}$ and
$0.00040 E_\text{H}$ for para- and ortho-H$_3^+$,
respectively \cite{Jaquet94jcp,Jaquet06}. These yield about
$0.020215 E_\text{H}$ for the average zero-point energy.
Note however, that the nuclear spins and zero point rotation
are not included in our model of H$_3^+$.

The lowest electronic excitation from the BO ground state is a direct
Franck--Condon one ($0.710
E_\text{H}$)\cite{Adamowicz09jcp1,Kreckel08jcp} to dissociative
potential curve: $\text{H}_3^+ \rightarrow \text{H}_2 + \text{H}^+$ or
$\text{H}_3^+ \rightarrow \text{H}_2^+ + \text{H}$.
\cite{Viegas07jcp,Adamowicz09jcp1} The dissociation energies ($D_e$)
are $0.169 E_\text{H}$ and $0.241 E_\text{H}$, respectively.

The linear geometry with equal bond lengths $1.53912a_0$ ($D_{\infty
h}$) is a saddle point on the BO PES at $-1.27868190 E_\text{H}$
\cite{Jaquet94jcp} or $0.06515 E_\text{H}$ above the BO energy at the
equilibrium geometry. This energy is usually called as the barrier to
linearity \cite{Oka03jcp}.  The zero Kelvin energetics is shown in
Fig.~\ref{Fig1} by the three horizontal lines.

\begin{figure}[t]
 \begin{center}
   \includegraphics{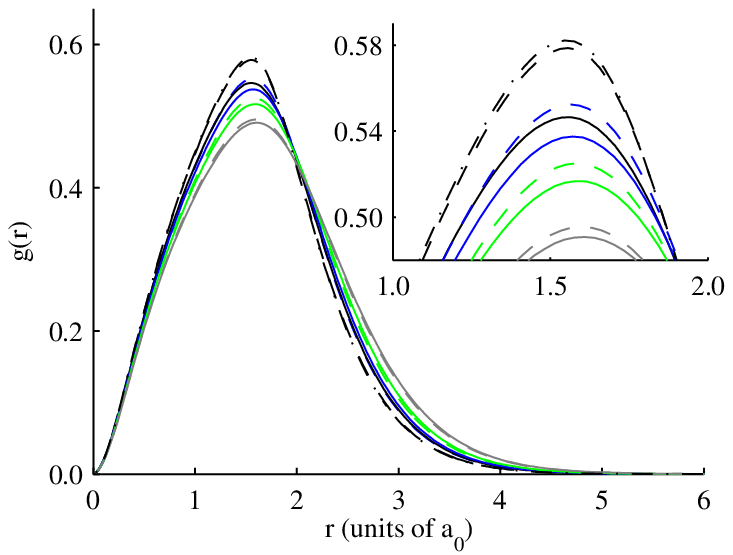}
   \caption{\label{Fig4}(Color online) Proton--electron pair
     correlation functions at the four temperatures from the full quantum
     statistical simulations, AQ (solid lines), and from simulations with the classical
     nuclei, CN (dashed lines). That from the BO scheme is
     given at the lowest (electronic) temperature, only (dash-dotted
     line). Notations are the same as in Fig.~\ref{Fig2}.}
\vskip1cm
   \includegraphics{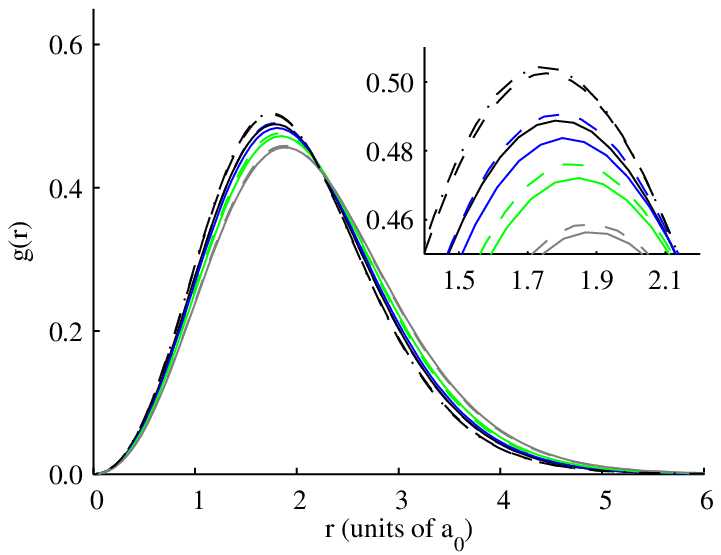}
   \caption{\label{Fig5}(Color online) Electron--electron pair
     correlation functions from the same simulations as those
     in Fig.~\ref{Fig4}.
     Notations are the same as in Figs.~\ref{Fig2} and \ref{Fig4}.}
 \end{center}
\end{figure}

\subsection{PIMC ground state: 160 K}

At our lowest simulation temperature, $T \approx 160$ K, the
electronic system is essentially in its ground state.
For the total energy we find $-1.3438(2) E_\text{H}$, see the
BO black triangles in Fig.~\ref{Fig1}. The thermal energy is
$k_BT = 0.000507 E_\text{H}$, and therefore, the contribution
from the rotational and vibrational excited states is also small
and we find  $-1.3406(29) E_\text{H}$,
see the CN red square in same Fig.
The full quantum simulation includes vibrational zero-point
contribution and yields $-1.3233(12) E_\text{H}$,
about $0.0205(14) E_\text{H}$ above the BO energy
in a good agreement with about $0.0202
E_\text{H}$ in Refs.~\cite{Jaquet94jcp,Jaquet06}.

From our AQ simulation we still find the equilateral triangle
configuration of the nuclei with the internuclear distances increased
to $\ka{R} = 1.723(4) a_0$, which indicates an increase of about
$0.073(4) a_0$, as compared with the zero Kelvin BO equilibrium
distance bond lengths.  Interestingly, within the error limits this is
the same as the bond length increase of the hydrogen molecule ion
H$_2^+$. The zero-point energy of H$_3^+$ is about $2.7$ times as
large as that of the H$_2^+$ ion \cite{Kylanpaa07}, as expected from
the increase of vibrational modes from one to three --- the zero-point
energy of our model does not contain the rotational zero-point energy,
as mentioned earlier.

The thermal motion (CN), alone, increases the bond length
to $\ka{R}= 1.658(4) a_0$, only, see the data in
Figs.~\ref{Fig2} and Fig.~\ref{Fig3}.
This clearly points out the difference between quantum and
thermal delocalization of nuclei at low $T$.

For the proton--electron and electron--electron interactions the
differences between our two approaches are smaller than in the
proton--proton case but still distinctive. Comparison of the fixed
nuclei simulation to the CN one shows that the two schemes give almost
identical distributions. The AQ distributions, however, cannot be
labeled identical with those from the CN or fixed nuclei
simulations. The distributions are given in Figs.~\ref{Fig4} and
\ref{Fig5}, where the notations are the same as in Fig.~\ref{Fig2}.

The calculations of the relativistic corrections involve, among
other things, evaluation of the contact densities,
$\ka{\delta(r_{ij})}$, for the electron--nuclei and the
electron--electron pairs \cite{Cencek98jcp}.
For the electron--nuclei contact density at the BO equilibrium
configuration we get  $0.1814(20)$, and
for the AQ case $0.1765(20)$. For the electron--electron pair we get
$0.0182(3)$ and $0.0166(3)$, for BO and AQ approaches
respectively.  The estimated uncertainties
due to extrapolation to the contact are given in parenthesis.
The zero Kelvin reference values
\cite{Cencek98jcp} for the BO case are $0.181242$ (electron--nuclei)
and $0.01838663$ (electron--electron). Thus, the quantum dynamics
of the nuclei turns out to be significant factor in lowering the contact
densities, too.

\begin{figure}[t]
 \begin{center}
   \includegraphics{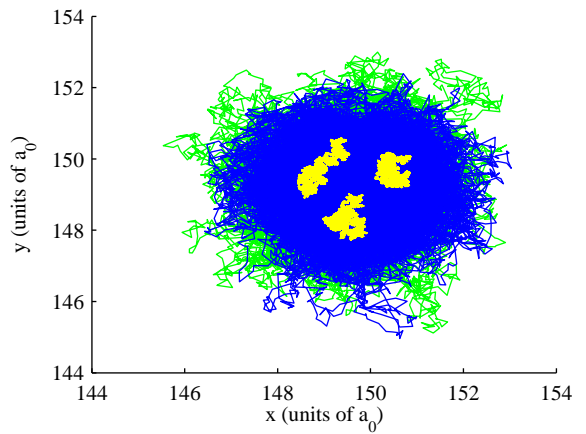}
   \caption{\label{Fig6}(Color online) xy-plane (z-projection)
     snapshot of the H$_3^+$ ion from quantum statistical simulation
     with Trotter number $2^{16}$, i.e.~temperature of about $160$ K,
     for all particles. ''Polymer rings'' describing the electrons
     are in the background (green and blue) and those of the nuclei
     are placed on top (yellow).}
 \end{center}
\end{figure}

See the snapshot of the AQ simulation in Fig.~\ref{Fig6} for some
intuition of the low-temperature quantum distributions in imaginary time.

\begin{table}[b]
\begin{center}
\caption{\label{Table1}Energetics of the H$_3^+$ molecular ion. The
energies are given in the units of Hartree (atomic units). Simulation
data is given with $2$SEM error estimates. BO refers to
Born--Oppenheimer calculation at equilibrium geometry. The reference
data is rounded to convenient accuracy.  The "barrier to linearity" is
$0.06515 E_\text{H} \approx 1.8$ eV above the $E_\text{BO}$ at $0$ K.}
\begin{tabular*}{0.48\textwidth}{l@{\hspace{0.5cm}}rccc}
\hline\hline
& \multicolumn{1}{c}{$T$} & $E_\text{BO}$ & $E_\text{CN}$ & $E_\text{AQ}$\\
\hline
Ref.~\cite{Adamowicz09jcp1} & $0$ K& $-1.343836$ & & $-1.323568$\footnotemark[1]\\
PIMC & $\sim 161$ K& $-1.3438(2)$ & $-1.3406(29)$ & $-1.3233(12)$\\
PIMC & $\sim 1365$ K& & $-1.3236(8)$ & $-1.3142(4)$\\
PIMC & $\sim 2570$ K& & $-1.3033(7)$ & $-1.2977(6)$\\
PIMC & $\sim 3856$ K& $-1.3438(2)$ & $-1.2810(8)$ & $-1.2770(2)$\\
PIMC & $\sim 3999$ K& & $-1.1469(9)$& $-1.2750(4)$\\
PIMC & $\sim 4050$ K& &  & $-1.2734(9)$\\
\hline\hline
\end{tabular*}
\end{center}
\footnotetext[1]{For ortho-H$_3^+$
estimated by using Refs.~ \cite{Jaquet06} and \cite{Adamowicz09jcp1}.}
\end{table}

\subsection{High temperature phenomena}

With the increasing temperature the increasing contribution from
rovibrational excitations is clearly seen in the total energies shown
in Fig.~\ref{Fig1}.  Contributions from the electronic excitations do
not appear, because the lowest excitation energy $0.710 E_\text{H}$ is
much too high as compared to the thermal energy $k_B T$.
Consequently, the equilibrium geometry BO energy depends on the
temperature almost negligibly.  For convenience, the essential
energetics related data has been collected into Table
\ref{Table1}, also.

As expected, the increase in the total energy due to the classical
rovibrational degrees of freedom is $9 \times \tfrac{1}{2} k_B T $,
defining the slope of the CN line.  The most prominent quantum feature
in AQ curve is, of course, the zero-point vibration energy.  At higher
temperatures, however, by comparing the AQ and CN curves we see that
the quantum nature of nuclear dynamics becomes less important, except
for dissociation.

At the dissociation limit we find the molecule with quantum nuclei
somewhat more stable than the one with classical nuclei.  With the
relatively low density, $(300 a_0)^3$, the molecule is mainly kept in
one piece above $4000$ K in the former case, whereas more dissociated
in the latter.  The total energy becomes higher for the CN than the AQ
case slightly below $4000$ K, see Table \ref{Table1}.  The total
energies at this crossing point are above the "barrier to
linearity",\cite{Jaquet94jcp,Oka03jcp} already.

At higher temperatures, $T\ge 4100$ K, other configurations,
such as $\text{H}_2 + \text{H}^+$, $\text{H}_2^+ + \text{H}$
and $2\text{H} + \text{H}^+$, start playing more significant
role in the equilibrium dissociation--recombination processes.
These will considered in our next study.

The nuclear pair correlation function or bond length distributions,
Figs.~\ref{Fig2} and \ref{Fig3}, follow the energetics discussed,
above.  There, the zero-point vibration in AQ case is seen even
better.  At the zero Kelvin limit both the expectation value and the
distribution, in particular, are significantly different from those of
the CN case.

The temperature dependence in the other pair correlation functions is
weak, see Figs.~\ref{Fig4} and \ref{Fig5}.  Obviously, this is the
case, because electrons do not present a quantum-to-classical
transition in the temperature range considered, now.  Thus, the
evolution in distributions in Figs.~\ref{Fig4} and \ref{Fig5}
following the rising temperature arises from the changes in the
nuclear dynamics, and mostly, from the change in the conformation or
the bond lengths, presented in Fig.~\ref{Fig3}.

\section{Conclusions}
In this study, the path integral Monte Carlo method was shown to be a
successful approach for examination of quantum statistics of the
five-particle molecule, H$_3^+$ ion.  The method is based on the
finite temperature mixed state description, and thus, it gives
information, which is complementary to the high-accuracy zero Kelvin
description of conventional quantum chemistry.  It was also shown how
contributions from quantum and thermal dynamics to particle
distributions and correlation functions can be sorted out, and
furthermore, quantum to classical dynamics transition can be
monitored.

Our approach is fully basis set and trial wave function free. It is
based on the Coulomb interactions, only, and allows the most
transparent interpretation of consequent particle--particle
correlations.

Simulation at $160$ K essentially reproduces the zero Kelvin data from
conventional quantum chemistry.  Of course, a proper extrapolation to
$0$ K can be done for more accuracy.  Born--Oppenheimer (BO) potential
energy surface and the equilibrium geometry can be found by using
classical nuclei with fixed coordinates. Description of the zero-point
motion within our nonadiabatic five-body quantum simulation gives the
vibrational zero-point energy accurately. We find an increase of
$0.073(4)a_0$ in the bond length due to the nonadiabatic zero-point
vibration.  The classical thermal contribution at $160$ K is
$0.008(4)a_0$, only.

With the raising temperature the rovibrational excitations contribute
to the energetics, as expected, whereas the electronic part remains in
its ground state in the spirit of BO approximation. At about $4000$ K
the H$_3^+$ ion dissociates, weakly depending on the ion density.  We
find that the full quantum molecule dissociates at slightly higher
temperature compared to the one, where the nuclei are modeled by
classical particles with thermal dynamics, only.  Thus, we conclude
the necessity of the quantum character of the protons in the correct
description of dissociation.

We find that the nuclear quantum dynamics has a distinctive effect on
the pair correlation functions, too.  This is least for the
electron--electron pair correlation function, stronger for the
electron--proton one and largely increased in the proton--proton
correlations. These are seen in the contact densities, and
consequently, in the relativistic corrections where relevant.

\section{Acknowlegements}

For financial support we thank the Academy of Finland, and for
computational resources the facilities of Finnish IT Center for
Science (CSC) and Material Sciences National Grid Infrastructure
(M-grid, akaatti). We also thank Kenneth Esler and Bryan Clark for
their advise concerning the pair approximation.


\begin{thebibliography}{28}
\expandafter\ifx\csname natexlab\endcsname\relax\def\natexlab#1{#1}\fi
\expandafter\ifx\csname bibnamefont\endcsname\relax
  \def\bibnamefont#1{#1}\fi
\expandafter\ifx\csname bibfnamefont\endcsname\relax
  \def\bibfnamefont#1{#1}\fi
\expandafter\ifx\csname citenamefont\endcsname\relax
  \def\citenamefont#1{#1}\fi
\expandafter\ifx\csname url\endcsname\relax
  \def\url#1{\texttt{#1}}\fi
\expandafter\ifx\csname urlprefix\endcsname\relax\def\urlprefix{URL }\fi
\providecommand{\bibinfo}[2]{#2}
\providecommand{\eprint}[2][]{\url{#2}}

\bibitem[{\citenamefont{Oka}(1992)}]{Oka92revmodphys}
\bibinfo{author}{\bibfnamefont{T.}~\bibnamefont{Oka}}, \bibinfo{journal}{Rev.
  Mod. Phys.} \textbf{\bibinfo{volume}{64}}, \bibinfo{pages}{1141}
  (\bibinfo{year}{1992}).

\bibitem[{\citenamefont{Gottfried et~al.}(2003)\citenamefont{Gottfried, McCall,
  and Oka}}]{Oka03jcp}
\bibinfo{author}{\bibfnamefont{J.~L.} \bibnamefont{Gottfried}},
  \bibinfo{author}{\bibfnamefont{B.~J.} \bibnamefont{McCall}},
  \bibnamefont{and} \bibinfo{author}{\bibfnamefont{T.}~\bibnamefont{Oka}},
  \bibinfo{journal}{J. Chem. Phys.} \textbf{\bibinfo{volume}{118}},
  \bibinfo{pages}{10890} (\bibinfo{year}{2003}).

\bibitem[{\citenamefont{Kutzelnigg and Jaquet}(2006)}]{Jaquet06}
\bibinfo{author}{\bibfnamefont{W.}~\bibnamefont{Kutzelnigg}} \bibnamefont{and}
  \bibinfo{author}{\bibfnamefont{R.}~\bibnamefont{Jaquet}},
  \bibinfo{journal}{Phil.~Trans.~R.~Soc.~A} \textbf{\bibinfo{volume}{364}},
  \bibinfo{pages}{2855} (\bibinfo{year}{2006}).

\bibitem[{\citenamefont{Pavanello and Adamowicz}(2009)}]{Adamowicz09jcp1}
\bibinfo{author}{\bibfnamefont{M.}~\bibnamefont{Pavanello}} \bibnamefont{and}
  \bibinfo{author}{\bibfnamefont{L.}~\bibnamefont{Adamowicz}},
  \bibinfo{journal}{J. Chem. Phys.} \textbf{\bibinfo{volume}{130}},
  \bibinfo{pages}{034104} (\bibinfo{year}{2009}).

\bibitem[{\citenamefont{Kreckel et~al.}(2008)\citenamefont{Kreckel, Bing,
  Reinhardt, Petrignani, Berg, and Wolf}}]{Kreckel08jcp}
\bibinfo{author}{\bibfnamefont{H.}~\bibnamefont{Kreckel}},
  \bibinfo{author}{\bibfnamefont{D.}~\bibnamefont{Bing}},
  \bibinfo{author}{\bibfnamefont{S.}~\bibnamefont{Reinhardt}},
  \bibinfo{author}{\bibfnamefont{A.}~\bibnamefont{Petrignani}},
  \bibinfo{author}{\bibfnamefont{M.}~\bibnamefont{Berg}}, \bibnamefont{and}
  \bibinfo{author}{\bibfnamefont{A.}~\bibnamefont{Wolf}}, \bibinfo{journal}{J.
  Chem. Phys.} \textbf{\bibinfo{volume}{129}}, \bibinfo{pages}{164312}
  (\bibinfo{year}{2008}).

\bibitem[{\citenamefont{Thomson}(1911)}]{Thomson11}
\bibinfo{author}{\bibfnamefont{J.~J.} \bibnamefont{Thomson}},
  \bibinfo{journal}{Philos. Mag.} \textbf{\bibinfo{volume}{21}},
  \bibinfo{pages}{225} (\bibinfo{year}{1911}).

\bibitem[{\citenamefont{Oka}(1980)}]{Oka80prl}
\bibinfo{author}{\bibfnamefont{T.}~\bibnamefont{Oka}}, \bibinfo{journal}{Phys.
  Rev. Lett.} \textbf{\bibinfo{volume}{45}}, \bibinfo{pages}{531}
  (\bibinfo{year}{1980}).

\bibitem[{\citenamefont{Lystrup et~al.}(2008)\citenamefont{Lystrup, Miller,
  Russo, R.~J.~Vervack, and Stallard}}]{Miller08ApJ}
\bibinfo{author}{\bibfnamefont{M.~B.} \bibnamefont{Lystrup}},
  \bibinfo{author}{\bibfnamefont{S.}~\bibnamefont{Miller}},
  \bibinfo{author}{\bibfnamefont{N.~D.} \bibnamefont{Russo}},
  \bibinfo{author}{\bibfnamefont{J.}~\bibnamefont{R.~J.~Vervack}},
  \bibnamefont{and} \bibinfo{author}{\bibfnamefont{T.}~\bibnamefont{Stallard}},
  \bibinfo{journal}{Astrophys. J.} \textbf{\bibinfo{volume}{677}},
  \bibinfo{pages}{790} (\bibinfo{year}{2008}).

\bibitem[{\citenamefont{Koskinen et~al.}(2009)\citenamefont{Koskinen, Aylward,
  and Miller}}]{Koskinen09ApJ}
\bibinfo{author}{\bibfnamefont{T.~T.} \bibnamefont{Koskinen}},
  \bibinfo{author}{\bibfnamefont{A.~D.} \bibnamefont{Aylward}},
  \bibnamefont{and} \bibinfo{author}{\bibfnamefont{S.}~\bibnamefont{Miller}},
  \bibinfo{journal}{Astrophys. J.} \textbf{\bibinfo{volume}{693}},
  \bibinfo{pages}{868} (\bibinfo{year}{2009}).

\bibitem[{\citenamefont{Meyer et~al.}(1986)\citenamefont{Meyer, Botschwina, and
  Burton}}]{Meyer86jcp}
\bibinfo{author}{\bibfnamefont{W.}~\bibnamefont{Meyer}},
  \bibinfo{author}{\bibfnamefont{P.}~\bibnamefont{Botschwina}},
  \bibnamefont{and} \bibinfo{author}{\bibfnamefont{P.}~\bibnamefont{Burton}},
  \bibinfo{journal}{J. Chem. Phys.} \textbf{\bibinfo{volume}{84}},
  \bibinfo{pages}{891} (\bibinfo{year}{1986}).

\bibitem[{\citenamefont{R\"{o}hse et~al.}(1994)\citenamefont{R\"{o}hse,
  Kutzelnigg, Jaquet, and Klopper}}]{Jaquet94jcp}
\bibinfo{author}{\bibfnamefont{R.}~\bibnamefont{R\"{o}hse}},
  \bibinfo{author}{\bibfnamefont{W.}~\bibnamefont{Kutzelnigg}},
  \bibinfo{author}{\bibfnamefont{R.}~\bibnamefont{Jaquet}}, \bibnamefont{and}
  \bibinfo{author}{\bibfnamefont{W.}~\bibnamefont{Klopper}},
  \bibinfo{journal}{J. Chem. Phys.} \textbf{\bibinfo{volume}{101}},
  \bibinfo{pages}{2231} (\bibinfo{year}{1994}).

\bibitem[{\citenamefont{Cencek et~al.}(1998)\citenamefont{Cencek, Rychlewski,
  Jaquet, and Kutzelnigg}}]{Cencek98jcp}
\bibinfo{author}{\bibfnamefont{W.}~\bibnamefont{Cencek}},
  \bibinfo{author}{\bibfnamefont{J.}~\bibnamefont{Rychlewski}},
  \bibinfo{author}{\bibfnamefont{R.}~\bibnamefont{Jaquet}}, \bibnamefont{and}
  \bibinfo{author}{\bibfnamefont{W.}~\bibnamefont{Kutzelnigg}},
  \bibinfo{journal}{J. Chem. Phys.} \textbf{\bibinfo{volume}{108}},
  \bibinfo{pages}{2831} (\bibinfo{year}{1998}).

\bibitem[{\citenamefont{Bachorz et~al.}(2009)\citenamefont{Bachorz, Cencek,
  Jaquet, and Komasa}}]{Komasa09jcp}
\bibinfo{author}{\bibfnamefont{R.~A.} \bibnamefont{Bachorz}},
  \bibinfo{author}{\bibfnamefont{W.}~\bibnamefont{Cencek}},
  \bibinfo{author}{\bibfnamefont{R.}~\bibnamefont{Jaquet}}, \bibnamefont{and}
  \bibinfo{author}{\bibfnamefont{J.}~\bibnamefont{Komasa}},
  \bibinfo{journal}{J. Chem. Phys.} \textbf{\bibinfo{volume}{131}},
  \bibinfo{pages}{024105} (\bibinfo{year}{2009}).

\bibitem[{\citenamefont{Li and Broughton}(1987)}]{Li87}
\bibinfo{author}{\bibfnamefont{X.-P.} \bibnamefont{Li}} \bibnamefont{and}
  \bibinfo{author}{\bibfnamefont{J.~Q.} \bibnamefont{Broughton}},
  \bibinfo{journal}{J. Chem. Phys} \textbf{\bibinfo{volume}{86}},
  \bibinfo{pages}{5094} (\bibinfo{year}{1987}).

\bibitem[{\citenamefont{Ceperley}(1995)}]{Ceperley95}
\bibinfo{author}{\bibfnamefont{D.~M.} \bibnamefont{Ceperley}},
  \bibinfo{journal}{Rev. Mod. Phys} \textbf{\bibinfo{volume}{67}},
  \bibinfo{pages}{279} (\bibinfo{year}{1995}).

\bibitem[{\citenamefont{Pierce and Manousakis}(1999)}]{Pierce99}
\bibinfo{author}{\bibfnamefont{M.}~\bibnamefont{Pierce}} \bibnamefont{and}
  \bibinfo{author}{\bibfnamefont{E.}~\bibnamefont{Manousakis}},
  \bibinfo{journal}{Phys. Rev. B} \textbf{\bibinfo{volume}{59}},
  \bibinfo{pages}{3802} (\bibinfo{year}{1999}).

\bibitem[{\citenamefont{Kwon and Whaley}(1999)}]{Kwon99}
\bibinfo{author}{\bibfnamefont{Y.}~\bibnamefont{Kwon}} \bibnamefont{and}
  \bibinfo{author}{\bibfnamefont{K.~B.} \bibnamefont{Whaley}},
  \bibinfo{journal}{Phys. Rev. Lett.} \textbf{\bibinfo{volume}{83}},
  \bibinfo{pages}{4108(4)} (\bibinfo{year}{1999}).

\bibitem[{\citenamefont{Knoll and Marx}(2000)}]{Knoll00}
\bibinfo{author}{\bibfnamefont{L.}~\bibnamefont{Knoll}} \bibnamefont{and}
  \bibinfo{author}{\bibfnamefont{D.}~\bibnamefont{Marx}},
  \bibinfo{journal}{Europ. Phys J. D} \textbf{\bibinfo{volume}{10}},
  \bibinfo{pages}{353} (\bibinfo{year}{2000}).

\bibitem[{\citenamefont{Cuervo and Roy}(2006)}]{Cuervo06}
\bibinfo{author}{\bibfnamefont{J.~E.} \bibnamefont{Cuervo}} \bibnamefont{and}
  \bibinfo{author}{\bibfnamefont{P.-N.} \bibnamefont{Roy}},
  \bibinfo{journal}{J. Chem. Phys.} \textbf{\bibinfo{volume}{125}},
  \bibinfo{pages}{124314} (\bibinfo{year}{2006}).

\bibitem[{\citenamefont{Kyl\"anp\"a\"a and Rantala}(2009)}]{Kylanpaa09pra}
\bibinfo{author}{\bibfnamefont{I.}~\bibnamefont{Kyl\"anp\"a\"a}}
  \bibnamefont{and} \bibinfo{author}{\bibfnamefont{T.~T.}
  \bibnamefont{Rantala}}, \bibinfo{journal}{Phys. Rev. A}
  \textbf{\bibinfo{volume}{80}}, \bibinfo{pages}{024504}
  (\bibinfo{year}{2009}).

\bibitem[{\citenamefont{Kyl\"anp\"a\"a
  et~al.}(2007)\citenamefont{Kyl\"anp\"a\"a, Leino, and Rantala}}]{Kylanpaa07}
\bibinfo{author}{\bibfnamefont{I.}~\bibnamefont{Kyl\"anp\"a\"a}},
  \bibinfo{author}{\bibfnamefont{M.}~\bibnamefont{Leino}}, \bibnamefont{and}
  \bibinfo{author}{\bibfnamefont{T.~T.} \bibnamefont{Rantala}},
  \bibinfo{journal}{Phys.~Rev.~A} \textbf{\bibinfo{volume}{76}},
  \bibinfo{pages}{052508(7)} (\bibinfo{year}{2007}).

\bibitem[{\citenamefont{Anderson}(1992)}]{Anderson92jcp}
\bibinfo{author}{\bibfnamefont{J.}~\bibnamefont{Anderson}},
  \bibinfo{journal}{J. Chem. Phys.} \textbf{\bibinfo{volume}{93}},
  \bibinfo{pages}{3702} (\bibinfo{year}{1992}).

\bibitem[{\citenamefont{Feynman}(1998)}]{Fey72}
\bibinfo{author}{\bibfnamefont{R.~P.} \bibnamefont{Feynman}},
  \emph{\bibinfo{title}{Statistical Mechanics}} (\bibinfo{publisher}{Perseus
  Books}, \bibinfo{year}{1998}).

\bibitem[{\citenamefont{Storer}(1968)}]{Storer68}
\bibinfo{author}{\bibfnamefont{R.~G.} \bibnamefont{Storer}},
  \bibinfo{journal}{J.~Math.~Phys.} \textbf{\bibinfo{volume}{9}},
  \bibinfo{pages}{964} (\bibinfo{year}{1968}).

\bibitem[{\citenamefont{Metropolis et~al.}(1953)\citenamefont{Metropolis,
  Rosenbluth, Rosenbluth, Teller, and Teller}}]{Metro53}
\bibinfo{author}{\bibfnamefont{N.}~\bibnamefont{Metropolis}},
  \bibinfo{author}{\bibfnamefont{A.~W.} \bibnamefont{Rosenbluth}},
  \bibinfo{author}{\bibfnamefont{M.~N.} \bibnamefont{Rosenbluth}},
  \bibinfo{author}{\bibfnamefont{A.~H.} \bibnamefont{Teller}},
  \bibnamefont{and} \bibinfo{author}{\bibfnamefont{E.}~\bibnamefont{Teller}},
  \bibinfo{journal}{J.~Chem.~Phys.} \textbf{\bibinfo{volume}{21}},
  \bibinfo{pages}{1087} (\bibinfo{year}{1953}).

\bibitem[{\citenamefont{Chakravarty et~al.}(1998)\citenamefont{Chakravarty,
  Gordillo, and Ceperley}}]{Chakravarty98}
\bibinfo{author}{\bibfnamefont{C.}~\bibnamefont{Chakravarty}},
  \bibinfo{author}{\bibfnamefont{M.~C.} \bibnamefont{Gordillo}},
  \bibnamefont{and} \bibinfo{author}{\bibfnamefont{D.~M.}
  \bibnamefont{Ceperley}}, \bibinfo{journal}{J. Chem. Phys.}
  \textbf{\bibinfo{volume}{109}}, \bibinfo{pages}{2123} (\bibinfo{year}{1998}).

\bibitem[{\citenamefont{Herman et~al.}(1982)\citenamefont{Herman, Bruskin, and
  Berne}}]{Herman82}
\bibinfo{author}{\bibfnamefont{M.~F.} \bibnamefont{Herman}},
  \bibinfo{author}{\bibfnamefont{E.~J.} \bibnamefont{Bruskin}},
  \bibnamefont{and} \bibinfo{author}{\bibfnamefont{B.~J.} \bibnamefont{Berne}},
  \bibinfo{journal}{J.~Chem.~Phys.} \textbf{\bibinfo{volume}{76}},
  \bibinfo{pages}{5150} (\bibinfo{year}{1982}).

\bibitem[{\citenamefont{Viegas et~al.}(2007)\citenamefont{Viegas, Alijah, and
  Varandas}}]{Viegas07jcp}
\bibinfo{author}{\bibfnamefont{L.~P.} \bibnamefont{Viegas}},
  \bibinfo{author}{\bibfnamefont{A.}~\bibnamefont{Alijah}}, \bibnamefont{and}
  \bibinfo{author}{\bibfnamefont{A.~J.~C.} \bibnamefont{Varandas}},
  \bibinfo{journal}{J. Chem. Phys.} \textbf{\bibinfo{volume}{126}},
  \bibinfo{pages}{074309} (\bibinfo{year}{2007}).

\end{thebibliography}
\end{document}